\documentclass[aps,prb,twocolumn,amsmath,amssymb,superscriptaddress,letterpaper,longbibliography,10pt]{revtex4-2}
\usepackage{times}
\usepackage{amsfonts}
\usepackage{mathrsfs}
\usepackage[version=3]{mhchem}
\usepackage{graphicx}
\usepackage{dcolumn}
\usepackage{bm}
\usepackage{color}

\usepackage[colorlinks,bookmarks=false,citecolor=blue,linkcolor=red,urlcolor=blue]{hyperref}

\def\be{\begin{equation}}       \def\ee{\end{equation}}
\def\bea{\begin{eqnarray}}      \def\eea{\end{eqnarray}}

\begin{document}
\title{Magnetic configurations and excitations in high-$T_{c}$ multilayer nickelates }

\author{Jun Zhan}
\affiliation{Beijing National Laboratory for Condensed Matter Physics and Institute of Physics, Chinese Academy of Sciences, Beijing 100190, China}
\affiliation{School of Physical Sciences, University of Chinese Academy of Sciences, Beijing 100190, China}

 \author{Xianxin Wu}\email{xxwu@itp.ac.cn}
 \affiliation{Institute of Theoretical Physics,
Chinese Academy of Sciences, Beijing 100190, China}

\author{Jiangping Hu}\email{jphu@iphy.ac.cn}
\affiliation{Beijing National Laboratory for Condensed Matter Physics and Institute of Physics, Chinese Academy of Sciences, Beijing 100190, China}
\affiliation{School of Physical Sciences, University of Chinese Academy of Sciences, Beijing 100190, China}
\affiliation{ New Cornerstone Science Laboratory, Institute of Physics, Chinese Academy of Sciences, Beijing 100190, China}

\begin{abstract}

We investigate the magnetic ground states and transverse spin excitations of bilayer and trilayer nickelates within a multi-orbital itinerant framework. 
For the bilayer system, although Hartree-Fock calculations slightly favor a double-stripe order, the calculated excitation spectrum of the single-stripe state, characterized by an anisotropic low-energy cone at $\bm{Q}_{\text{BL}}$ and isotropic high-energy excitations near $\Gamma$, exhibits good qualitative agreement with recent RIXS and neutron scattering experiments.
We further identify mirror-even optical interlayer modes at $\bm{Q}_{\text{BL}}$ whose energies match the mirror-odd modes at $\Gamma$. For the trilayer system, both mirror-odd and mirror-even spin-density-wave states can be stabilized near $\bm{Q}_{\text{TL}}$, with the mirror-odd state lower in energy in the parameter regime studied. The mirror-odd state hosts an additional nearly gapless mode dominated by the middle layer, while the mirror-even state contains only one acoustic branch together with two gapped optical modes. Comparison with available RIXS data favors the mirror-odd spin-density-wave  scenario. Our results show that magnetic excitations provide a sensitive probe of the magnetic order and support a common itinerant origin of magnetism in multilayer nickelates.

\end{abstract}

\maketitle

\section{Introduction}

Recently, high-temperature ($T_c$) superconductivity was observed in the bilayer ($\text{La}_3\text{Ni}_2\text{O}_7$, LNO) and trilayer ($\text{La}_4\text{Ni}_3\text{O}_{10}$) nickelates of the Ruddlesden--Popper phase under pressure \cite{sun2023, JZhao2023T}, which has drawn tremendous research interest. Notably, the bilayer system with partial rare-earth element substitution exhibits an extraordinarily high transition temperature of approximately 100 K \cite{Qiu2025NdLNO}. In both systems, superconductivity emerges abruptly following a structural transition from the \textit{Amam}/\textit{P2$_1$/a} phase to the \textit{Fmmm}/\textit{I4mmm} phase as pressure increases. Distinct from the $d^9$ electronic configuration of cuprates and infinite-layer nickelates, these multilayer nickelates possess nominal Ni $d^{7.5}$ and $d^{7.4}$ configurations. Both $d_{x^2-y^2}$ and $d_{z^2}$ orbitals contribute to the low-energy electronic structure, owing to the presence of apical oxygens between the $\text{NiO}_2$ layers \cite{YaoDX, YZhang2023, Lechermann2023, Hirofumi2023possible, XWu, XJZhou2023, HHWen2023}. Remarkably, recent studies have also identified signatures of superconductivity in compressively strained LNO thin films grown on $\text{SrLaAlO}_4$ substrates, with $T_c$ exceeding 40 K at ambient pressure \cite{ko2024signatures, zhou2024ambientpressuresuperconductivityonset40, liu2025superconductivitynormalstatetransportcompressively, bhatt2025resolvingstructuraloriginssuperconductivity}. In both bulk and thin-film forms, superconductivity in these nickelates exists in close competition with spin density wave (SDW) orders, typically emerging only upon the suppression of the SDW order~\cite{sun2023, HQYuan2023, HHWen2023, chen2024electronic, ren_resolving_2025, XHChen2025-SDW}.

The pairing mechanism and symmetry of these materials are currently under intensive theoretical investigation, though a consensus has yet to be reached~\cite{Wang327prb, lu2024interlayer,HYZhangtype2,XWu,FangYang327prl,WeiLi327prl,Hirofumi2023possible,YifengYang327prb,YifengYang327prb2,YiZhuangYouSMG,tian2023correlation,Dagotto327prb,zhang2024structural,Jiang_2024,Lechermann2023,liao2023electron,ryee2024quenched,luo2023hightc,fan2023superconductivity,KuWeiprl,zhan2024cooperation,ChenHH2025,PhysRevB.111.144514,GuanGJ2025,WangQH2025_prl,Wang4310,Yang4310}. 
Theoretical calculations suggest that spin fluctuations tend to promote $s_{\pm}$-wave interlayer pairing \cite{Hirofumi2023possible, XWu, Wang327prb, FangYang327prl}, while complementary studies emphasize the role of interlayer exchange, Hund's coupling, and orbital hybridization in generating interlayer pairing within the strong-coupling limit \cite{lu2024interlayer, HYZhangtype2, WeiLi327prl}. However, the origin of the SDW and its relationship with superconductivity remain less theoretical explored~\cite{LeCC2025}. 
Drawing a parallel to the cuprates and iron-based superconductors, both magnetic order and superconductivity likely stem from a common root in electronic correlations and can be understood within a unified theoretical framework.
As the SDW phase sits in close proximity to superconductivity in the phase diagrams of both bilayer and trilayer systems~\cite{XHChen2025-SDW,YangJ2026TrilayerDW}, elucidating the origin of magnetism is vital to resolve the broader superconducting mechanism.

Experimentally, magnetic order and excitations in these systems have been probed by resonant inelastic x-ray scattering (RIXS)~\cite{chen2024electronic,chen2026,chan2026}, muon spin rotation ($\mu$SR)~\cite{ChenKW2024PRLSDW,Khasanov2025NatPhysDW}, and neutron scattering~\cite{Plokhikh2025NeutronSDW,ChenLX2026NeutronMagnetism}. Under ambient conditions, bilayer nickelates exhibit SDW order below 150 K with an in-plane ordering vector of $\bm{Q}_{\text{BL}}=(0.5, 0.5)\pi$ (tetragonal notation). While initial RIXS measurements of spin waves were interpreted as consistent with either single-stripe (with spinless sites) or double-stripe scenarios~\cite{chen2024electronic}, recent neutron scattering data suggest the single-stripe scenario is more probable~\cite{ChenLX2026NeutronMagnetism}. In the trilayer system, the magnetic order intertwined with charge order occurs below 140 K and exhibits an incommensurate in-plane wave vector of $\bm{Q}_{\text{TL}}=(0.62, 0.62)\pi$~\cite{zhang2020intertwined}. It has been suggested that this order displays a unique layer-dependent pattern: the top and bottom layers are antiferromagnetically coupled, while the middle layer possesses a nearly vanishing magnetic moment. 
Despite these observations, the underlying mechanism driving these SDW states in both systems remains unclear. From a symmetry perspective, the SDW order in both bilayer and trilayer nickelates exhibit mirror-odd parity and similar wave vectors, suggesting a common electronic origin~\cite{LeCC2025}. Notably, these magnetic states cannot be naturally explained by a localized Heisenberg model. In such models, a strong interlayer exchange typically leads to interlayer spin singlets, which would preclude the formation of the observed in-plane magnetic order and fail to account for the vanishing moment in the middle layer of the trilayer system. In contrast, within an itinerant formalism, mirror-parity-selective scattering may play an essential role in driving magnetic order~\cite{LeCC2025}, as evidenced by the pocket-dependent SDW gaps observed in trilayer systems~\cite{YangJ2026TrilayerDW}. 
With available experimental data on magnetic excitations for both systems, 
it is essential to evaluate whether a unified itinerant framework can account for both the ground states and their excitations, helping to resolve the ongoing debate regarding the true magnetic structure.

In this work, we investigate the magnetic ground states and transverse spin excitations of bilayer and trilayer nickelates within a multi-orbital itinerant framework. For bilayer nickelates, our Hartree-Fock calculations indicate that the double-stripe order is slightly energetically favored over the single-stripe order. However, the calculated excitation spectrum of the SS state, characterized by an anisotropic low-energy cone at $\bm{Q}_{\text{BL}}$ and isotropic high-energy excitations around $\Gamma$, shows good qualitative agreement with recent RIXS and neutron scattering experiments. Furthermore, we identify optical interlayer modes in the mirror-even channel at $\bm{Q}_{\mathrm{BL}}$ for both configurations, with energies identical to those of the mirror-odd modes at $\Gamma$.
For trilayer nickelates, our self-consistent calculations indicate that both mirror-odd and mirror-even SDW states can be stabilized near $\bm{Q}_{\mathrm{TL}}$, with the mirror-odd phase emerging as the energetically favorable ground state. This mirror-odd SDW shares several characteristic features with the bilayer system but also hosts an additional, nearly gapless mode dominated by the middle layer. In contrast, the mirror-even SDW displays a single gapless mode accompanied by two gapped optical modes at $\bm{Q}_{\mathrm{TL}}$. Comparison with available RIXS data indicates that the mirror-odd SDW scenario is more consistent with experimental observations.
Our results demonstrate that the magnetic states in multilayer nickelates are well-captured within an itinerant framework and magnetic excitations can serve as a sensitive probe for identifying the nature of magnetic order.

\section{Model and formalism}

We consider a multiorbital Hamiltonian consisting of a tight-binding part and onsite interactions,
\begin{equation}
    \mathcal{H}=\mathcal{H}_{0}+\mathcal{H}_{\mathrm{int}}.
\end{equation}
The low-energy degrees of freedom are the Ni $e_g$ orbitals on each NiO$_2$ layer.
We denote the layer by $\ell=1,\cdots,N_L$, the orbital by $\mu,\nu=x,z$, with $x=d_{x^2-y^2}$ and $z=d_{z^2}$, and the spin by $\sigma$.
Combining layer and orbital labels into a flavor index $a=(\ell,\mu)$, the kinetic Hamiltonian is written as
\begin{equation}
    \mathcal{H}_{0}
    =
    \sum_{\bm{k}\sigma}
    c^{\dagger}_{a\bm{k}\sigma}
    \left[h_{0}(\bm{k})\right]_{ab}
    c_{b\bm{k}\sigma},
    \label{eq:h0_orbital}
\end{equation}
where repeated flavor indices are summed.
For the bilayer we use the tight-binding model of Ref.~\cite{WuRaghu2026}, while for the trilayer we use the ARPES-fitted model of Ref.~\cite{YangJ2026TrilayerDW}.
These models retain the relevant low-energy bands near the Fermi level.

In both systems, the layer degree of freedom allows the electronic states to be classified approximately by mirror parity.
In the bilayer, $c_{\pm,\mu}=(c_{1\mu}\pm c_{2\mu})/\sqrt{2}$ gives mirror-even and mirror-odd sectors.
The SDW instability considered below is associated primarily with scattering between the mirror-even bonding $\alpha$ sheet and the mirror-odd antibonding $\beta$ sheet at
$\bm{Q}_{\mathrm{BL}}\approx (\pi/2,\pi/2)$.
In the trilayer, the outer-layer combinations $(c_{1\mu}\pm c_{3\mu})/\sqrt{2}$ together with the middle-layer orbital $c_{2\mu}$ produce bonding, nonbonding, and antibonding branches.
The dominant magnetic channel connects the mirror-even bonding $\alpha$ band and the mirror-odd nonbonding $\beta'$ band near $\bm{Q}_{\mathrm{TL}}\simeq (0.62\pi,0.62\pi)$.
In this itinerant perspective, the magnetic instability is selected by Fermi-surface scattering between bands with opposite mirror paritiy~\cite{LeCC2025,YangJ2026TrilayerDW}.

The interaction part is the onsite Hubbard--Kanamori interaction,
\begin{equation}
    \begin{aligned}
    &\mathcal{H}_{\mathrm{int}}= \sum_{i\ell\mu}Un_{i\ell\mu\uparrow}n_{i\ell\mu\downarrow}+\sum_{i\ell,\mu\neq\nu}J_Pc_{i\ell\mu\uparrow}^{\dagger}c_{i\ell\mu\downarrow}^{\dagger}c_{i\ell\nu\downarrow}c_{i\ell\nu\uparrow}\\
    &+\sum_{i\ell,\mu<\nu,\sigma\sigma'}(U^{\prime}n_{i\ell\mu\sigma}n_{i\ell\nu\sigma^{\prime}}+J_Hc_{i\ell\mu\sigma}^{\dagger}c_{i\ell\nu\sigma}c_{i\ell\nu\sigma^{\prime}}^{\dagger}c_{i\ell\mu\sigma^{\prime}}),
    \end{aligned}
\end{equation}
Here $U$ and $U'$ are the intra- and inter-orbital Coulomb repulsions, $J_H$ is Hund's coupling, and $J_P$ is pair hopping.
We use the standard Kanamori relations $U=U'+2J_H$ and $J_P=J_H$~\cite{Kanamori}.
In the transverse spin channel, the corresponding onsite interaction matrix in the orbital particle-hole basis $(xx,zx,xz,zz)$ is
\begin{equation}
	\hat{V}_{\mathrm{s}}=
	\begin{pmatrix}
		U & 0 & 0 & J_H\\
		0 & U' & J_P & 0\\
		0 & J_P & U' & 0\\
		J_H & 0 & 0 & U
	\end{pmatrix}.
	\label{eq:kanamori_vertex}
\end{equation}

\subsection{Hartree--Fock method for SDW}

For a commensurate SDW with $N_Q\bm{Q}$ equal to a reciprocal lattice vector, we introduce the folded spinor
\begin{equation}
    \Psi_{\bm{k}\sigma}
    =
    \left(
    c_{a,\bm{k},\sigma},
    c_{a,\bm{k}+\bm{Q},\sigma},
    \cdots,
    c_{a,\bm{k}+(N_Q-1)\bm{Q},\sigma}
    \right)^T,
    \label{eq:folded_spinor}
\end{equation}
where $\bm{k}$ lies in the magnetic Brillouin zone and the flavor index $a=(\ell,\mu)$ is implicit in each block.
The spin-density matrix with transfer momentum $m\bm{Q}$ is defined as
\begin{equation}
    \Phi^{(m)}_{ab}
    =
    \frac{1}{2N}
    \sum_{\bm{k}r\sigma}
    \sigma
    \left\langle
    c^{\dagger}_{a,\bm{k}+(r+m)\bm{Q},\sigma}
    c_{b,\bm{k}+r\bm{Q},\sigma}
    \right\rangle .
    \label{eq:orbital_order}
\end{equation}
The factor $1/2$ makes $\Phi^{(m)}$ the spin density rather than the density difference.
The diagonal components describe the onsite spin density of a given layer and orbital, while the off-diagonal components describe interorbital or interlayer particle-hole coherence.

The bilinear Hartree--Fock Hamiltonian in the folded basis is
\begin{equation}
    \mathcal{H}_{\mathrm{HF}}
    =
    \sum_{\bm{k}\sigma}
    \Psi_{\bm{k}\sigma}^{\dagger}
    \hat{H}^{\mathrm{HF}}_{\sigma}(\bm{k})
    \Psi_{\bm{k}\sigma}
    .
\end{equation}
with matrix elements
\begin{equation}
    \left[\hat{H}^{\mathrm{HF}}_{\sigma}(\bm{k})\right]_{ar,br'}
    =
    \left[h_0(\bm{k}+r\bm{Q})\right]_{ab}\delta_{rr'}
    +
    \sigma \Delta^{(r-r')}_{ab}.
    \label{eq:hf_matrix}
\end{equation}
The constant terms generated by the mean-field decoupling are included when comparing Hartree--Fock energies, but do not enter the quasiparticle spectrum or the transverse susceptibility.
For the onsite SDW states considered in this work, the mean field is diagonal in the layer-orbital flavor,
\begin{equation}
    \Delta^{(m)}_{(\ell\mu)(\ell'\nu)}
    =
    \delta_{\ell\ell'}\delta_{\mu\nu}M^{(m)}_{\ell\mu},
    \label{eq:onsite_sdw_field}
\end{equation}
and the self-consistent equations are
\begin{equation}
\begin{aligned}
    M^{(m)}_{\ell x}
    &=
    -U\Phi^{(m)}_{(\ell x)(\ell x)}
    -J_H\Phi^{(m)}_{(\ell z)(\ell z)},\\
    M^{(m)}_{\ell z}
    &=
    -U\Phi^{(m)}_{(\ell z)(\ell z)}
    -J_H\Phi^{(m)}_{(\ell x)(\ell x)}.
\end{aligned}
    \label{eq:orbital_hf_equations}
\end{equation}
Although the self-consistent fields in Eq.~(\ref{eq:orbital_hf_equations}) are local, the eigenstates of Eq.~(\ref{eq:hf_matrix}) inherit the full layer and orbital structure of $h_0(\bm{k})$.
Consequently, the SDW gap carries nontrivial layer and orbital form factors.
These form factors determine the mirror parity of the ordered state, the layer distribution of the ordered moment, and the symmetry channels in which the spin spectral weight appears.

For the bilayer, the relevant order has $N_Q=4$ and is dominated by components that, after transformation to the mirror-band basis, connect the bonding $\alpha$ and antibonding $\beta$ sectors.
The relative phase between the $+\bm{Q}_{\mathrm{BL}}$ and $-\bm{Q}_{\mathrm{BL}}$ components controls the real-space spin texture and distinguishes the spin-spinless single stripe from double-stripe-like configurations.
For the trilayer, the same orbital-basis construction is applied to the $\alpha$-$\beta'$ and $\alpha$-$\beta$ channels.
We use a commensurate approximant to the experimental incommensurate vector $\bm{Q}_{\mathrm{TL}}$; the specific choice is stated in the corresponding Hartree--Fock and RPA calculations below.

\subsection{Transverse spin susceptibility}

The magnetic excitation spectrum is obtained from the transverse spin response in the self-consistent SDW state~\cite{Brydon2009,Knolle2011Multiorbital}.
For orbital/layer flavors $a,b,c,d$, we first resolve the physical spin-flip operator into orbital/layer channels,
\begin{equation}
    S^{+}_{ab}(\bm{q})
    =
    \sum_{\bm{k}}
    c^{\dagger}_{a,\bm{k}+\bm{q},\uparrow}
    c_{b,\bm{k},\downarrow},
\label{eq:spin_flip_bilinear}
\end{equation}
and define
\begin{equation}
    \chi^{+-}_{ab,cd}(\bm{q},\bm{q}',\tau)
    =
    \left\langle
    T_{\tau}S^{+}_{ab}(\bm{q},\tau)
    S^{-}_{cd}(-\bm{q}',0)
    \right\rangle .
\label{eq:chi_tensor_def}
\end{equation}
In the paramagnetic state the two external momenta are identical.
In the SDW state the magnetic order acts as a static Bragg field, so a probe at $\bm{q}$ also couples to $\bm{q}+\lambda\bm{Q}$.
Thus the susceptibility is a matrix in the combined index
$I=(\lambda,ab)$, with $\lambda=0,\cdots,N_Q-1$.
For a period-four stripe, for instance, the same response contains the four transfer sectors
$\bm{q}$, $\bm{q}+\bm{Q}$, $\bm{q}+2\bm{Q}$, and $\bm{q}+3\bm{Q}$.

The bare susceptibility $\hat{\chi}^{0,+-}$ is the particle-hole bubble built from the HF quasiparticles.
Let $|n,\bm{k},\sigma\rangle$ be an eigenstate of Eq.~(\ref{eq:hf_matrix}) in the folded basis.
The spin-flip vertex in channel $I=(\lambda,ab)$ is the matrix element
\begin{equation}
    \Gamma^{I}_{mn}(\bm{k},\bm{q})
    =
    \left\langle
    m,\bm{k}+\bm{q},\uparrow
    \left|
    S^{+}_{ab}(\bm{q}+\lambda\bm{Q})
    \right|
    n,\bm{k},\downarrow
    \right\rangle .
    \label{eq:sdw_vertex}
\end{equation}
This SDW-state vertex matrix is the coherence factor of the ordered state.
It encodes the layer, orbital, and folded-momentum content of the initial and final quasiparticles, and therefore controls the distribution of spectral weight among different symmetry channels.
With this notation,
\begin{equation}
    \chi^{0,+-}_{I J}(\bm{q},\omega)
    =
    -\frac{1}{N}
    \sum_{\bm{k}mn}
    \Gamma^{I}_{mn}
    \Gamma^{J*}_{mn}
    \frac{
        f(E_{n\downarrow})-f(E_{m\uparrow})
    }{
        \omega+i\eta+E_{n\downarrow}-E_{m\uparrow}
    },
    \label{eq:chi0}
\end{equation}
where the energies and vertices are evaluated for the transition
$|n,\bm{k},\downarrow\rangle\rightarrow |m,\bm{k}+\bm{q},\uparrow\rangle$.

The RPA susceptibility is obtained by summing the ladder series in the transverse spin channel,
\begin{equation}
    \chi^{+-}_{\mathrm{RPA}}(\bm{q},\omega)
    =
    \left[
    \hat{\bm{1}}-\chi^{0,+-}(\bm{q},\omega)\hat{V}_{\mathrm{SDW}}
    \right]^{-1}
    \chi^{0,+-}(\bm{q},\omega).
    \label{eq:rpa}
\end{equation}
The interaction vertex has the same channel structure as $\hat{\chi}^{0,+-}$.
Since the Hubbard-Kanamori interaction is local on each Ni site, the pair-space vertex is block diagonal in the layer index,
\begin{equation}
    \hat{V}_{\mathrm{pair}}
    =
    \mathrm{diag}
    \left(
        \hat{V}_{\mathrm{s}}^{(1)},
        \hat{V}_{\mathrm{s}}^{(2)},
        \cdots,
        \hat{V}_{\mathrm{s}}^{(N_L)}
    \right),
    \qquad
    \hat{V}_{\mathrm{s}}^{(\ell)}=\hat{V}_{\mathrm{s}},
    \label{eq:pair_vertex}
\end{equation}
where $\hat{V}_{\mathrm{s}}$ is given in Eq.~(\ref{eq:kanamori_vertex}) in the onsite orbital-pair basis.
The SDW folding simply repeats this local vertex in each transfer sector,
\begin{equation}
    \hat{V}_{\mathrm{SDW}}
    =
    \hat{1}_{N_Q}\otimes \hat{V}_{\mathrm{pair}} .
    \label{eq:sdw_rpa_vertex}
\end{equation}
The off-diagonal coupling between different transfer sectors is then generated by the ordered-state bubble itself, through the SDW quasiparticle vertices in Eq.~(\ref{eq:sdw_vertex}).

Finally, we project the RPA tensor onto physical spin operators with definite layer character.
Since spin probes couple to the onsite spin density, the projection is taken over the intraorbital spin channels.
For a layer-orbital form factor $W$ acting within this intraorbital subspace,
\begin{equation}
    S^{+}_{W}(\bm{q})=
    \sum_{\bm{k}}
    c^{\dagger}_{a,\bm{k}+\bm{q},\uparrow}
    W_{ab}
    c_{b,\bm{k},\downarrow},
\end{equation}
the corresponding response is
\begin{equation}
    \chi^{+-}_{W}(\bm{q},\omega)
    =
    W^{*}_{I}
    \chi^{+-}_{\mathrm{RPA},IJ}(\bm{q},\omega)
    W_{J}.
    \label{eq:physical_projection}
\end{equation}
Here $I,J$ are restricted to the intraorbital components of the orbital-pair and folded-momentum channel space.
The imaginary part of $\chi^{+-}_{W}$ gives the spin spectral weight directly measured, up to standard magnetic form factors and polarization factors, in inelastic neutron scattering.
For the bilayer, the layer-even and layer-odd spin channels are represented by
\begin{equation}
    W^{\mathrm{BL}}_{\mathrm{even}}
    =
    \tau_{0}\otimes\sigma_{0},
    \qquad
    W^{\mathrm{BL}}_{\mathrm{odd}}
    =
    \tau_{z}\otimes\sigma_{0},
    \label{eq:bilayer_form_factors}
\end{equation}
where $\tau_i$ and $\sigma_i$ act in the layer and orbital spaces, respectively.
The layer-odd channel becomes interband in the bonding-antibonding basis and therefore directly probes the opposite-mirror-parity spin response~\cite{LeCC2025}.
For orbital-resolved spectra, $\sigma_0$ is replaced by the corresponding orbital projector.

For the trilayer, we decompose the response into the outer-layer mirror-odd, outer-layer mirror-even, and middle-layer channels,
\begin{equation}
\begin{aligned}
    W^{\mathrm{TL}}_{\mathrm{odd}}
    &=
    P_{\mathrm{odd}}\otimes\sigma_{0},
    \qquad
    P_{\mathrm{odd}}=\mathrm{diag}(1,0,-1),
    \\
    W^{\mathrm{TL}}_{\mathrm{outer}}
    &=
    P_{\mathrm{outer}}\otimes\sigma_{0},
    \qquad
    P_{\mathrm{outer}}=\mathrm{diag}(1,0,1),
    \\
    W^{\mathrm{TL}}_{\mathrm{middle}}
    &=
    P_{\mathrm{middle}}\otimes\sigma_{0},
    \qquad
    P_{\mathrm{middle}}=\mathrm{diag}(0,1,0).
\end{aligned}
    \label{eq:trilayer_form_factors}
\end{equation}
This decomposition is useful for the mirror-odd SDW, whose ordered moment resides mainly on the two outer layers with opposite signs.
For the mirror-even SDW, it is more natural to use the layer form factors
\begin{equation}
\begin{aligned}
    W^{\mathrm{TL}}_{\mathrm{ME},G}
    &=
    P_{\mathrm{ME},G}\otimes\sigma_{0},
    \qquad
    P_{\mathrm{ME},G}=\frac{1}{2}\mathrm{diag}(1,-\sqrt{2},1),
    \\
    W^{\mathrm{TL}}_{\mathrm{ME},O}
    &=
    P_{\mathrm{ME},O}\otimes\sigma_{0},
    \qquad
    P_{\mathrm{ME},O}=\frac{1}{2}\mathrm{diag}(1,\sqrt{2},1).
\end{aligned}
    \label{eq:trilayer_mirror_even_form_factors}
\end{equation}
The first follows the relative phase of the mirror-even ordered moment and carries the acoustic response, whereas the second describes the corresponding optical fluctuation.
Together with the mirror-odd form factor in Eq.~(\ref{eq:trilayer_form_factors}), these projections separate the layer character of the trilayer spin excitations.

\begin{figure}[t]
	\centerline{\includegraphics[width=0.5\textwidth]{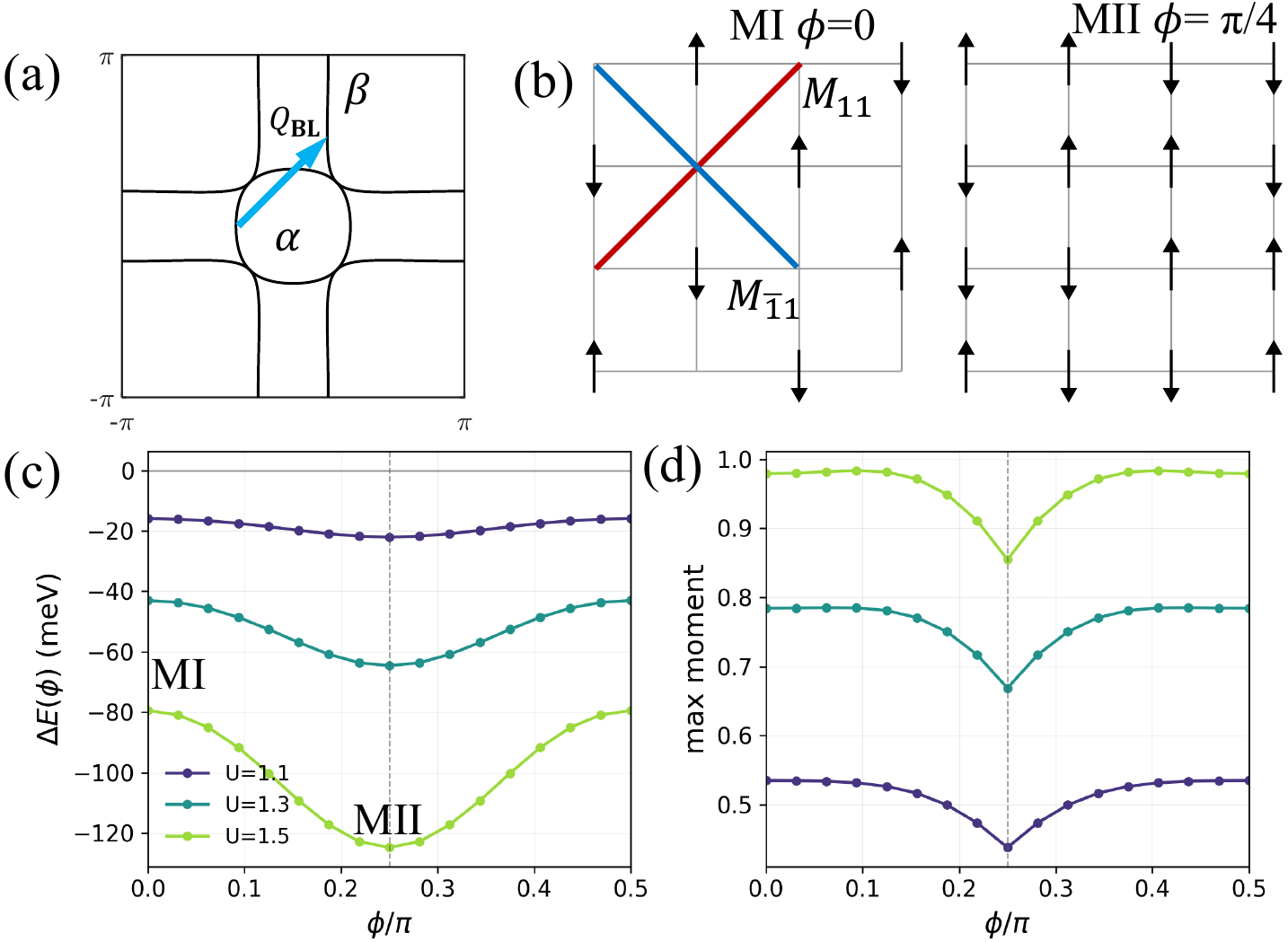}}
	\caption{Fermi surfaces, magnetic configurations, and energetics of bilayer nickelates. (a) Fermi surfaces. (b) Real-space patterns of the single-stripe ($\phi=0$) and double-stripe ($\phi=\pi/4$) magnetic orders, where sites without arrows are spinless. (c) The energy per site with respect to the normal state and (d) maximum real-space magnetic moment as functions of the phase parameter $\phi$.  \label{fig:bilayer_spectrum}}
\end{figure}

\begin{figure*}[t]
	\centerline{\includegraphics[width=0.99\textwidth]{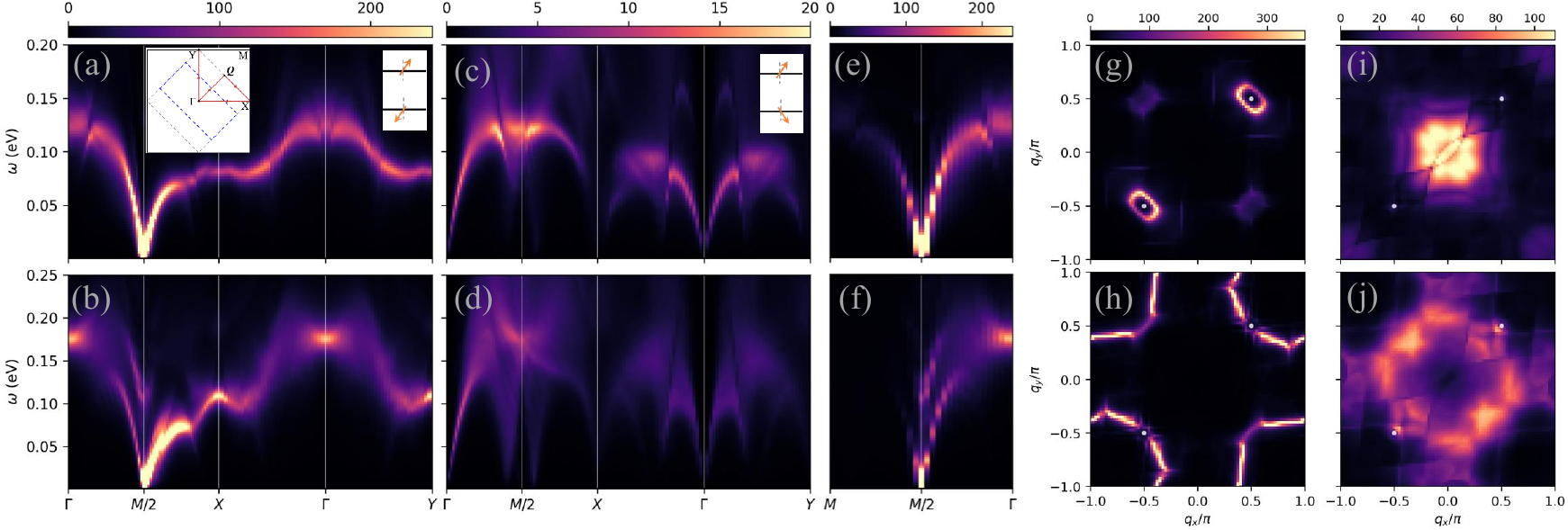}}
\caption{
Transverse spin excitations of the $U=1.3$ eV Hartree--Fock states.
The inset in (a) shows the unfolded Brillouin zone, the magnetic Brillouin zone associated with the period-four ordering vector $Q=(\pi/2,\pi/2)$, and the momentum paths used for the spectra.
(a,b) Mirror-odd spectra of the single-stripe and double-stripe states, respectively.
(c,d) Corresponding mirror-even spectra.
The small insets in (a) and (c) illustrate the bilayer mirror-odd and mirror-even modes.
(e,f) Mirror-odd spectra along the $M \to \Gamma$ path for the single-stripe and double-stripe states.
(g,h) Fixed-energy mirror-odd momentum maps at $\omega=0.05$ eV for the single-stripe and double-stripe states.
(i,j) Corresponding maps at $\omega=0.13$ eV.
\label{fig:bilayer_spectrum2}}
\end{figure*}

\section{Magnetic state and excitations of bilayer nickelate}

We first apply the above framework to the bilayer nickelate.
The normal-state Fermiology contains a bonding $\alpha$ pocket and an antibonding $\beta$ pocket of opposite mirror parity, as shown in Fig.~\ref{fig:bilayer_spectrum}(a).
The dominant opposite-mirror-parity scattering between $\alpha$ and $\beta$ with a vector $\bm{Q}_{\mathrm{BL}}\approx(\pi/2,\pi/2)$ is crucial for both magnetic order and superconducting pairing~\cite{LeCC2025}.
Here we study the magnetic state selected when this scattering channel condenses in the particle-hole sector.
Within the Hartree-Fock calculations, the magnetic moment is modeled using the ansatz:
\begin{equation}
\bm{M}^l(\bm{R}) = M^l_0 \hat{\bm{z}} \cos(\bm{Q}_{\text{BL}} \cdot \bm{R} + \phi),
\end{equation}
where $l=\{t,b\}$ denotes the top and bottom layers. In our calculations, the onsite interaction together with the mirror-odd band form factor favors an interlayer antiferromagnetic relation, $M^t_0=-M^b_0$, consistent with scattering between the $\alpha$ and $\beta$ pockets~\cite{LeCC2025}. In this parametrization, $\phi=0$ corresponds to a single-stripe magnetic order with spinless sites, while $\phi=\pi/4$ gives the conventional double-stripe order with nearly uniform moment amplitude. The real-space patterns are illustrated in Fig.~\ref{fig:bilayer_spectrum}(b).
Symmetry analysis reveals distinct features for these two states: the single-stripe order ($\phi = 0$) preserves a mirror symmetry along the diagonal $[11]$ direction passing through the spinless sites (red line in Fig.~\ref{fig:bilayer_spectrum}(b)), an effective mirror symmetry along the $[1\bar{1}]$ direction when combined with a spin-flip operation (blue line in Fig.~\ref{fig:bilayer_spectrum}(b)), and the $PT$ (Parity times Time reversal) symmetry at the spinless site.  These symmetries are absent in the double-stripe configuration and the last symmetry has a significant effect on the magnetic excitations.

Figure~\ref{fig:bilayer_spectrum}(c) shows the calculated energy of the magnetic states as a function of the phase $\phi$. Our results indicate that the double-stripe magnetic order is the energetically favorable ground state, with the energy difference between the single-stripe and double-stripe phases increasing as the interaction strengths $U$ and $J$ increase. Furthermore, the maximum real-space magnetic moment as a function of $\phi$ is shown in Fig.~\ref{fig:bilayer_spectrum}(d). We find that the maximum moment decreases continuously as $\phi$ evolves from $0$ to $\pi/4$. In the absence of charge degrees of freedom, this preference for the double-stripe phase is physically reasonable: the double-stripe order facilitates the opening of larger magnetic gaps across the Fermi surface, thereby maximizing the condensation energy and lowering the total electronic energy of the system.

\begin{figure*}[t]
	\centerline{\includegraphics[width=0.98\textwidth]{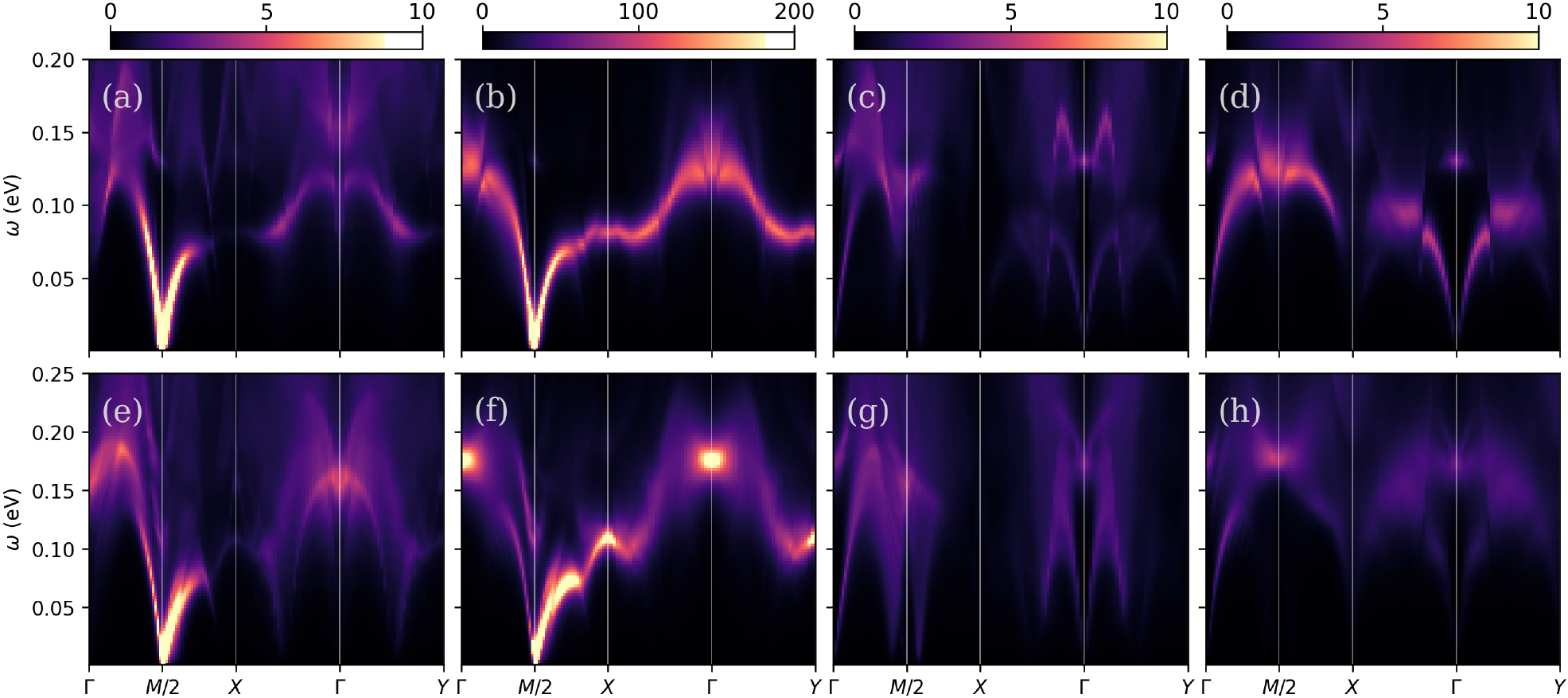}}
	\caption{
Orbital-resolved transverse spin excitation spectra for the $U=1.3$ eV Hartree--Fock states along the high-symmetry momentum path.
The upper row shows the single-stripe state, while the lower row shows the double-stripe state.
Panels (a,e) show the $d_{x^2-y^2}$-orbital-resolved mirror-odd channel, panels (b,f) the $d_{z^2}$-orbital-resolved mirror-odd channel, panels (c,g) the $x$-orbital-resolved mirror-even channel, and panels (d,h) the $z$-orbital-resolved mirror-even channel.
Here $M/2=Q=(\pi/2,\pi/2)$ is the period-four ordering wave vector.
\label{fig:bilayer_orbital_spectrum}}
\end{figure*}

Fig.~\ref{fig:bilayer_spectrum2} displays the imaginary part of the transverse spin susceptibility $\hat{\chi}^{+-}_{\mathrm{RPA}}(\bm{q},\omega)$ with $U=1.3$ eV and $J/U=0.1$ along a high-symmetry path for the two representative magnetic states. In the mirror-odd channel, both magnetic states feature a linearly dispersing gapless mode around $\bm{Q}_{\mathrm{BL}}$ [see Figs.~\ref{fig:bilayer_spectrum2}(a) and \ref{fig:bilayer_spectrum2}(b)], consistent with the Goldstone mode associated with broken spin-rotation symmetry. These cone are degenerate due to the PT symmetry with the single-stripe (SS) phase. 
In the double-stripe (DS) state, an additional intense feature bends downward in energy along the $\bm{Q}_{\mathrm{BL}}$--X direction. This feature does not correspond to a collective spin wave, but instead arises from spin-flip particle-hole excitations between remnant parts of pockets.
Notably, the spin-wave cones are highly anisotropic along the two orthogonal directions around $\bm{Q}_{\mathrm{BL}}$, reflecting  the itinerant character of the magnetic excitations around $\bm{Q}_{\mathrm{BL}}$ and $C_4$ symmetry breaking of these magnetic states. The spin-wave velocity along $\Gamma-\bm{Q}_{\mathrm{BL}}$ is higher than in the orthogonal direction. Furthermore, this velocity in the DS state is greater than that in the SS state. 
The velocity ratio $v_{11}/v_{1\bar{1}}$ is approximately 1.4 for the single-stripe (SS) state and 1.2 for the double-stripe (DS) state.
In addition, the DS state exhibits a weak gapped excitation at $\bm{Q}_{\mathrm{BL}}$, with enhanced spectral weight along the $[11]$ direction. This is attributed to strong cone splitting stemming from the broken PT symmetry within the DS state.
At higher energies, magnetic excitations are also visible along the $\Gamma-X/Y$ paths and exhibit a localized suppression of spectral weight around $\Gamma$, which is likely attributed to particle-hole excitations. Unlike the low-energy dispersion near $\bm{Q}_{\mathrm{BL}}$, the spectra along these two orthogonal directions are nearly identical for both magnetic states.
In both cases, a slight softening is observed near the magnetic Brillouin zone boundary (close to the X point), which we identify as a saddle point. In the DS state, the spectrum near $\Gamma$ becomes  broadened, which we attribute to damping from particle-hole excitations.

In the mirror-even channel [Figs.~\ref{fig:bilayer_spectrum2}(c) and \ref{fig:bilayer_spectrum2}(d)], both magnetic states display a gapped optical mode at $\bm{Q}_{\mathrm{BL}}$. The energy of this optical mode coincides with that of the $\Gamma$-point mode in the mirror-odd channel. A weak gapless feature is also present near the $\Gamma$ point.
The orbital-resolved transverse susceptibilities in Fig.~\ref{fig:bilayer_orbital_spectrum} show that the mirror-odd spin wave is dominated by the intraorbital $d_{z^2}$ contribution, while the $d_{x^2-y^2}$ component also contributes finite spectral weight to the Goldstone mode. By contrast, the optical modes in the mirror-even channel receive contributions from both orbitals.

Moreover, the momentum-space distribution of spectral weight further distinguishes the two magnetic states [Figs.~\ref{fig:bilayer_spectrum2}(e) and \ref{fig:bilayer_spectrum2}(f)]. Along M--$\Gamma$, the spin-wave intensity in the SS state is nearly symmetric with respect to $\bm{Q}_{\mathrm{BL}}$, whereas the DS state shows a pronounced asymmetry. This difference follows from the symmetry distinction discussed above: the SS state preserves the $M_{11}$ mirror symmetry, while the DS state does not.
These dispersions demonstrate that the gapless cones in the SS state are degenerate. In contrast, the DS state features both a gapless cone and a gapped branch near the ordering vector $\bm{Q}_{\mathrm{BL}}$ due to the broken PT symmetry.
The corresponding constant-energy intensity maps of the mirror-odd transverse susceptibility at $\omega=0.05$ eV and $0.13$ eV are shown in Figs.~\ref{fig:bilayer_spectrum2}(g-j). At $\omega=0.05$ eV, the elliptical ring around $\bm{Q}_{\mathrm{BL}}$ in the SS state directly visualizes the anisotropic spin-wave cone, while the arc-like feature in the DS state reflects the asymmetric spectral weight around $\bm{Q}_{\mathrm{BL}}$. At $\omega=0.13$ eV, both states exhibit an approximately isotropic ring around $\Gamma$, whereas the DS state additionally shows a weak ring-like feature around $\bm{Q}_{\mathrm{BL}}$.

\section{Magnetic state and excitations of trilayer nickelate}

\begin{figure}[t]
	\centerline{\includegraphics[width=0.50\textwidth]{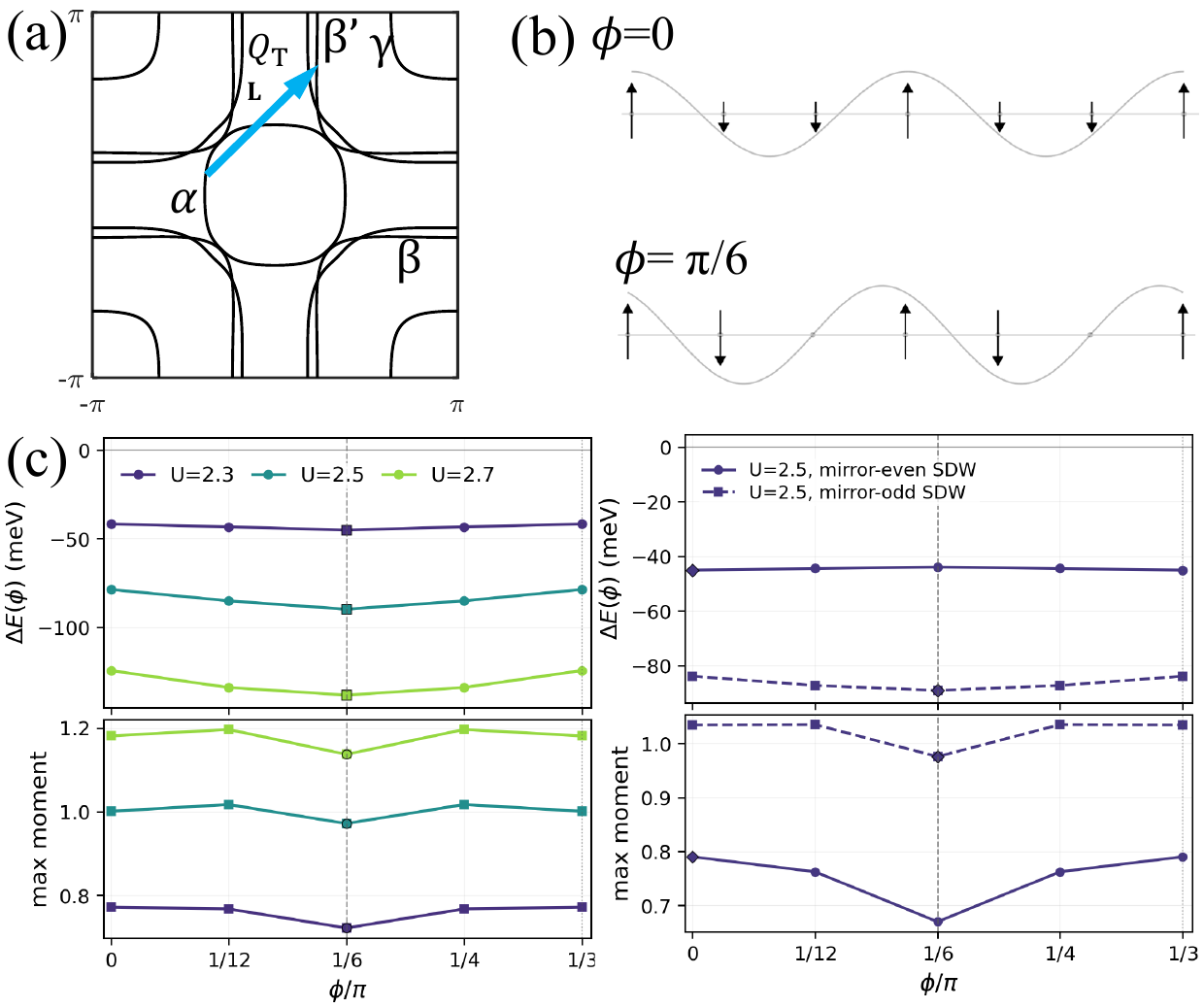}}
\caption{
(a) Fermi surface for the opposite mirror-parity nesting case, where the mirror-even $\alpha$ pocket and mirror-odd $\beta'$ pocket are connected by the nesting vector $\mathbf{Q}_{\rm TL}$.
(b) Schematic real-space spin modulations for two representative SDW phases, $\phi=0$ and $\phi=\pi/6$.
(c) The energy per site with respect to the normal state, $\Delta E(\phi)$, and the maximum real-space magnetic moment for the mirror-odd SDW driven by opposite mirror-parity nesting. The minimum occurs at $\phi=\pi/6$ for $U=2.3$, $2.5$, and $2.7$ eV.
(d) Corresponding phase dependence for the same mirror-parity nesting case at $U=2.5$ eV, comparing mirror-even and mirror-odd SDW solutions. The mirror-even solution is favored at $\phi=0$ modulo the period-three translation symmetry, while the mirror-odd branch has its minimum near $\phi=\pi/6$.
\label{fig4}}
\end{figure}

\begin{figure}[t]
	\centerline{\includegraphics[width=0.50\textwidth]{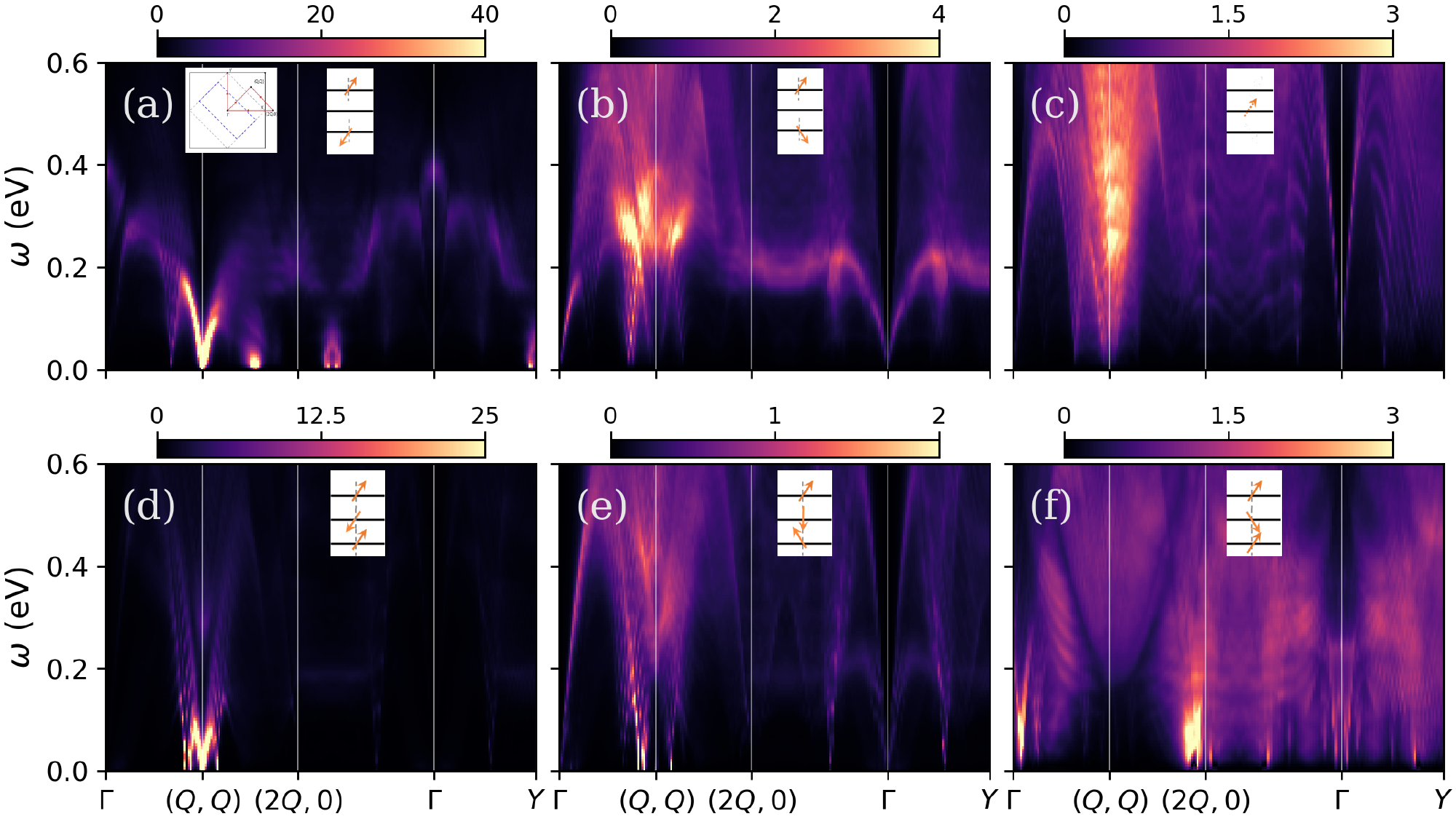}}
\caption{
Layer-resolved transverse spin spectra of the trilayer nickelate model. The spectra are evaluated along the high-symmetry path $\Gamma \rightarrow (Q,Q) \rightarrow (2Q,0) \rightarrow \Gamma \rightarrow Y$, with $Q=2\pi/3$. Panels (a)--(c) show the mirror-odd SDW state driven by opposite-mirror-parity nesting between the mirror-even $\alpha$ pocket and the mirror-odd $\beta'$ pocket at $U=2.3$ eV, projected onto the mirror-odd $(1,0,-1)/\sqrt{2}$, outer-layer mirror-even $(1,0,1)/\sqrt{2}$, and middle-layer $(0,1,0)$ form-factor channels, respectively. Panels (d)--(f) show the mirror-even SDW state driven by same-mirror-parity nesting between the mirror-even $\alpha$ and $\beta$ pockets at $U=2.5$ eV, projected onto the mirror-even Goldstone $(1,-\sqrt{2},1)/2$, mirror-even optical $(1,\sqrt{2},1)/2$, and mirror-odd $(1,0,-1)/\sqrt{2}$ channels, respectively.
The insets illustrate the corresponding  mirror-odd and mirror-even modes.
In the opposite-parity case, the dominant low-energy Goldstone branch appears in the mirror-odd channel, whereas in the same-parity case the low-energy Goldstone response is concentrated in the mirror-even Goldstone channel, with the remaining channels carrying optical spectral weight.
\label{fig:mu0p111_mirrorodd_spectrum}}
\end{figure}

We next perform Hartree--Fock calculations for the trilayer nickelate, using a tight-binding model fitted to the ARPES band structure.
As shown in Fig.~\ref{fig4}(a), the normal-state Fermiology consists of four Fermi pockets: two mirror-even bonding $\alpha$ and $\gamma$ pockets, a mirror-even antibonding $\beta$ pocket, and a mirror-odd nonbonding $\beta'$ pocket.
The $\alpha$ pocket is nested with both the $\beta$ and $\beta'$ pockets, with close nesting wave vectors.
In particular, scattering between the $\alpha$ and $\beta'$ pockets connects states of opposite mirror parity and therefore favors a mirror-odd SDW state.
By contrast, scattering between the $\alpha$ and $\beta$ pockets occurs within the same mirror sector and favors a mirror-even SDW state with antiferromagnetic alignment between adjacent layers.
The ordering wave vector in the mirror-odd channel is consistent with neutron scattering measurements.
Since the precise magnetic configuration has not yet been fully established experimentally, we consider SDW ansatzes in both the mirror-odd and mirror-even channels and compare their magnetic excitations.

Similar to the bilayer case, we describe the magnetic moment using the ansatz
\begin{equation}
\bm{M}^l(\bm{R}) =
M^l_0 \hat{\bm{z}}
\cos(\tilde{\bm{Q}}_{\text{TL}}\cdot \bm{R}+\phi),
\end{equation}
where $l=\{t,m,b\}$ labels the top, middle, and bottom layers.
For the Hartree--Fock energetics, we use the commensurate wave vector $\tilde{\bm{Q}}=(2\pi/3,2\pi/3)$
as a period-three approximant to the experimental ordering vector $\bm{Q}_{\text{TL}}\simeq(0.62\pi,0.62\pi)$.
In the mirror-odd channel, the magnetic moments satisfy $M^t_0=-M^b_0$ and $M^m_0=0$, so that the middle layer carries no static moment.
In the mirror-even channel, the outer layers are aligned with each other, $M^t_0=M^b_0$, while the middle layer is antiferromagnetically aligned with them, $\mathrm{sgn}(M^m_0)=-\mathrm{sgn}(M^{t/b}_0)$.
As in the bilayer case, the phase $\phi$ controls the in-plane distribution of the magnetic moment.

Figures~\ref{fig4}(c) and \ref{fig4}(d) show the Hartree--Fock phase dependence of the trilayer SDW states.
For the mirror-odd state, the energy minimum occurs at $\phi=\pi/6$ for $U=2.3$, $2.5$, and $2.7$ eV.
At this phase, the period-three modulation contains nearly spinless sites, similar to the single-stripe texture in the bilayer system.
The accompanying real-space moment follows the same phase dependence, indicating that the condensation energy is controlled by both the magnetic amplitude and the layer-mirror form factor.
Figure~\ref{fig4}(d) compares the mirror-odd and mirror-even solutions at $U=2.5$ eV.
The mirror-even branch is minimized at $\phi=0$ modulo period-three translations, whereas the mirror-odd branch remains lowest near $\phi=\pi/6$.
In the parameter range shown in Fig.~\ref{fig4}, the mirror-odd SDW has a lower Hartree--Fock energy than the mirror-even SDW, supporting the opposite-mirror-parity $\alpha$-$\beta'$ scattering mechanism~\cite{YangJ2026TrilayerDW}.

Figures~\ref{fig4}(c) and \ref{fig4}(d) present the Hartree-Fock energies and magnetic moments as functions of the phase $\phi$ for the mirror-odd and mirror-even channels, respectively. With the default Fermiology shown in Fig~\ref{fig4}(a), the mirror-odd SDW is the only robust solution, with an energy minimum at $\phi=\pi/6$ corresponding to a single-stripe order with spinless sites. To study the properties of mirror-even SDW, we slightly shifted the Fermi level to enhance the nesting in the mirror-even channel and the results are in shown in Fig.~\ref{fig4}(d). While this shift allows both the mirror-odd and mirror-even SDW states to be stabilized as self-consistent solutions, the mirror-odd state remains  energetically more favorable. In this regime, the mirror-even SDW favors a configuration at $\phi=0$ and the maximum real-space magnetic moment decreases continuously as $\phi$ evolves from $0$ to $\pi/6$ [Fig.~\ref{fig4}(d)]. These results suggest the robustness of the mirror-odd magnetic ground state, supporting the opposite-mirror-parity $\alpha$-$\beta'$ scattering mechanism~\cite{YangJ2026TrilayerDW}.

We further study the corresponding magnetic excitations of these mirror-odd and mirror-even magnetic orders to identify their characteristics.
Figure~\ref{fig:mu0p111_mirrorodd_spectrum} shows the imaginary part of
$\hat{\chi}^{+-}_{\mathrm{RPA}}(\bm{q},\omega)$ along a high-symmetry momentum path.
Panels (a)--(c) show the mirror-odd SDW state at $U=2.3$ eV, while panels (d)--(f) show the mirror-even SDW state at $U=2.5$ eV.
In both calculations we take $J/U=0.1$.
In the mirror-odd SDW state, the two outer layers are antiferromagnetically coupled, while the middle layer carries no static ordered moment. As a result, the middle layer has no local SDW exchange field and remains only dynamically coupled to the ordered outer layers. This effectively partitions the system into a bilayer and an independent monolayer, generating a characteristic layer-selective excitation structure.
The outer-layer mirror-odd channel exhibits a linearly dispersing spin wave around $\bar{\bm{Q}}_{\text{TL}}$.
This mode extends toward the $\Gamma$ point at higher energy, although its spectral intensity is reduced.
The mirror-even excitation channel involving the outer layers contains a gapped optical mode at $\bar{\bm{Q}}_{\text{TL}}$, whose energy matches that of the $\Gamma$-point mode in the mirror-odd channel. These features are consistent with an effective bilayer behavior, dicussed above.
Interestingly, the middle layer also contributes an almost gapless excitation near $\bar{\bm{Q}}_{\text{TL}}$, which is particularly prominent near $\Gamma$ via SDW folding due to the absence of masking signals from other branches.
This soft mode is not a conventional spin wave of a statically ordered moment; rather, it reflects the strongly enhanced transverse response of a layer that is coupled to the ordered outer layers but is not itself exchange pinned.

By contrast, the mirror-even SDW state exhibits a different excitation structure because the middle layer carries a finite ordered moment antiferromagnetically aligned with the outer layers.
In this state, the middle layer feels a nonzero molecular field from the ordered background, so its transverse fluctuation is exchange pinned and does not produce an additional nearly gapless mode.
In the mirror-even response channel, a linearly dispersing spin wave emerges around $\bar{\bm{Q}}_{\text{TL}}$ and extends toward the $\Gamma$ point.
In the outer-layer mirror-odd channel, a gapped optical mode appears at intermediate energy at $\bar{\bm{Q}}_{\text{TL}}$ and a weak gapless mode around $\Gamma$.
In addition, a second optical mode in the mirror-even channel associated with relative fluctuations between adjacent layers appears at higher energy with strongly damped spectral weight, which can be inferred from the weak intensity around $\bar{\bm{Q}}_{\text{TL}}$ shown in Fig.~\ref{fig:mu0p111_mirrorodd_spectrum}(f).
Thus, the mirror-even SDW state contains one gapless mode and two optical modes near $\bar{\bm{Q}}_{\text{TL}}$.
In contrast, the mirror-odd SDW state contains two nearly gapless modes and one optical mode.
These qualitative differences in the spin excitation spectra provide a direct way to distinguish the two candidate magnetic states experimentally.

\section{Discussion and conclusion}

Our results show that different magnetic orders give rise to distinct magnetic excitation spectra in multilayer high-$T_c$ nickelates, providing a useful way to identify the underlying magnetic state experimentally.
While the energy scales of magnetic excitations in our Hartree-Fock calculations are significantly higher than the values observed in experiments, a common feature of mean-field approaches, the calculated excitation spectra are qualitatively consistent with experimental observations.

In bilayer nickelates, the DS order exhibits an additional high-energy excitation at $\bm{Q}_{\mathrm{BL}}$ (the energy is comparable to that at the X point) and a gapless mode along $\Gamma$--M with strongly asymmetric spectral intensity.
These qualitative features are not consistent with recent RIXS and neutron scattering measurements.
By contrast, the SS state produces an elliptical low-energy ring excitation around $\bm{Q}_{\mathrm{BL}}$ and an approximately isotropic high-energy ring excitation around $\Gamma$, in qualitative agreement with the main features observed by neutron scattering. The observed slight gap feature around 25 meV at $\bm{Q}_{\mathrm{BL}}$ can be captured in a magnetic configuration with a small $\phi\neq0$, where the spinless site have a small moment. This slightly breaks the PT symmetry and leads to the slight gapped feature.
Although our calculation does not produce a low-energy spin gap at $\bm{Q}_{\mathrm{BL}}$, this discrepancy may originate from the absence of single-ion anisotropy in the present model.
Overall, the excitation spectrum supports the SS order as the magnetic state realized in bilayer nickelates.
At the same time, the Hartree--Fock energy of the SS state is higher than that of the DS state, suggesting that additional charge degrees of freedom or lattice effects may be important for stabilizing the experimentally relevant SS configuration.

In trilayer nickelates, the mirror-odd and mirror-even SDW states exhibit qualitatively different magnetic excitations because of their distinct interlayer magnetic structures.
The mirror-odd SDW has the layer pattern $(M,0,-M)$, so that the middle layer carries no static moment and is not pinned by a static exchange field.
As a result, its spectrum contains one Goldstone mode, one nearly gapless middle-layer-dominated mode, and one gapped optical mode near $\bm{Q}_{\mathrm{TL}}$.
By contrast, the mirror-even SDW has the layer pattern $(M,-M_m,M)$, with a finite middle-layer moment antiferromagnetically coupled to the outer layers.
The middle-layer fluctuation is then exchange pinned, and the spectrum contains only one Goldstone mode together with two gapped optical modes.
Available RIXS measurements appear more consistent with the mirror-odd SDW, for which only one optical mode is clearly identified.
At ambient pressure, the mirror symmetry is weakly broken, which may induce a small but finite moment on the middle layer even in the mirror-odd SDW state.
Such a moment would harden the middle-layer-dominated soft mode near $\bm{Q}_{\mathrm{TL}}$.

The high-energy optical spin-wave modes are a characteristic consequence of strong interlayer coupling in multilayer nickelates.
Experimentally, these modes may be isolated by tuning the out-of-plane momentum $L$, which controls the interlayer interference factor~\cite{YBCOoptical}.
Similar optical modes have been detected at low energies in cuprates~\cite{YBCOoptical,BSCCO}, but in multilayer nickelates they are expected to appear at higher energies because of the stronger interlayer hybridization and magnetic coupling.
The optical-mode gap therefore provides a direct measure of the effective interlayer magnetic coupling.

In summary, we have studied the magnetic ground states and transverse spin excitations of bilayer and trilayer nickelates within a multi-orbital itinerant framework.
For bilayer nickelates, we find that the DS order is lower in Hartree--Fock energy than the SS order with spinless sites.
However, the calculated excitation spectrum of the SS state, including an anisotropic low-energy cone around $\bm{Q}_{\mathrm{BL}}$ and an isotropic high-energy excitation around $\Gamma$, is more consistent with recent RIXS and neutron scattering experiments.
This suggests that the experimentally realized SDW in bilayer nickelates is likely the SS order, although additional interactions beyond the present Hartree--Fock treatment may be needed to stabilize it energetically.
We also find optical interlayer modes at $\bm{Q}_{\mathrm{BL}}$ in the mirror-even channel for both bilayer magnetic states.
For trilayer nickelates, both mirror-odd and mirror-even SDW states appear near $\bm{Q}_{\mathrm{TL}}$ in our calculations, with the mirror-odd SDW having lower energy.
The mirror-odd SDW shares several features with the bilayer case but also hosts an additional nearly gapless middle-layer-dominated mode, reflecting the absence of a static moment on the middle layer.
In contrast, the mirror-even SDW exhibits one gapless mode and two gapped optical modes near $\bm{Q}_{\mathrm{TL}}$.
The available RIXS data seem more compatible with the mirror-odd SDW~\cite{chen2026,chan2026}.
These results demonstrate that magnetic excitation spectra provide a sensitive probe of the magnetic order in multilayer nickelates.
Together with the itinerant pairing scenario driven by scattering between Fermi pockets of opposite mirror parity~\cite{LeCC2025,YangJ2026TrilayerDW}, our work supports a unified itinerant description in which both SDW order and superconductivity arise from the same underlying low-energy electronic structure.

{\it Acknowledgments}. We acknowledge the supports by the Ministry of Science and Technology (Grant No. 2022YFA1403901), National Natural Science Foundation of China (No. 11920101005, No. 11888101, No. 12047503, No. 12322405, No. 12104450) and the New Cornerstone Investigator Program. X.W. is supported by the National Key R\&D Program of China (Grant No. 2023YFA1407300) and the National Natural Science Foundation of China (Grants No. 12574151, 12447103 and 12447101).

{\it Note added}. 
During the preparation of this work, we became aware of a related study on magnetic excitations in nickelates using a band-basis formalism~\cite{mei2026itinerantnaturespindensitywaveorder}. In contrast, our study employs a multi-orbital formalism that explicitly retains the orbital and layer structure of the ordered states. This approach allows us to investigate competing magnetic configurations and identify their unique characteristics. Specifically, our identification of distinct acoustic and optical spin-excitation branches near the ordering vectors in bilayer and trilayer systems provides a theoretical basis for distinguishing between potential magnetic ground states in experiment.


%

\end{document}